\documentstyle[12pt]{article}
\topmargin .cm
\newcommand{\nn}{\nonumber}
\newcommand{\bd}{\begin{document}}
\newcommand{\ed}{\end{document}}
\newcommand{\bc}{\begin{center}}
\newcommand{\ec}{\end{center}}
\newcommand{\be}{\begin{eqnarray}}
\newcommand{\ee}{\end{eqnarray}}
\renewcommand{\thefootnote}{\alph{footnote}}
\newcommand{\se}{\section}
\newcommand{\sse}{\subsection}
\newcommand{\bi}{\bibitem}
\newcommand{\text}{\rm}
\newcommand{\bflr}{\begin{flushright}}
\newcommand{\eflr}{\end{flushright}}
\def\figcap{\section*{Figure Captions\markboth
     {FIGURECAPTIONS}{FIGURECAPTIONS}}\list
     {Figure \arabic{enumi}:\hfill}{\settowidth\labelwidth{Figure 999:}
     \leftmargin\labelwidth
     \advance\leftmargin\labelsep\usecounter{enumi}}}
\let\endfigcap\endlist \relax
\input psfig

\begin{document}

\begin{titlepage}
 \vskip 1.cm
 \null
\begin{center}
 \vspace{.15in}
{\Large {\bf
CP Violation in the Decay $\eta\to \pi^+\pi^-\gamma$
}}\\
\vspace{1.0cm}  \par
 \vskip 2.1em
 {\large
  \begin{tabular}[t]{c}
{\bf  C.~Q.~Geng$^{a,b}$, J.~N.~Ng$^b$, and T.~H.~Wu$^{a,b}$}
\\
\\
{\sl ${}^a$Department of Physics, National Tsing Hua University}
\\  {\sl  $\ $ Hsinchu, Taiwan, Republic of China }
\\
\\
{\sl ${}^b$Theory Group, TRIUMF, 4004 Wesbrook Mall}
\\ {\sl $\ $ Vancouver, B.C. V6T 2A3, Canada}
   \end{tabular}}
 \par \vskip 5.3em

 {\Large\bf Abstract}
\end{center}

We study the CP violating effects in the decay of
$\eta\to\pi^+\pi^- \gamma$.
We show that to have CP violation in the decay, one has to consider
both linear and circular photon polarizations.
In the standard model, the polarizations are vanishingly small.
However, model-independently, i.e. using only experimental constraint 
imposed by the limit on $Br(\eta\to\pi^+\pi^-)$ it can be up to $O(10\%)$.
We also explore various possible operators and we find that
the tensor type operator, possibly arising from a nonzero CP
violating electric dipole moment of the strange quark, can induce a 
sizable linear photon polarization.


\end{titlepage}


Both CP violation (CPV) and time-reversal (T) 
violation (TV) have been measured experimentally in the $K^0$ system 
\cite{KCPV}. More recently, CPV has also been seen in another 
flavor-changing  $B$-meson decays \cite{BCPV}.
However, the origin of the violations remains unclear. In the standard 
model, CPV or TV
arises from a unique physical phase in the Cabibbo-Kobayashi-Maskawa (CKM)
quark mixing matrix \cite{ckm}. It is also intimately 
connected to flavor changing processes. To ensure that this phase is indeed the
source of CP  or T violation, and to gain a deeper understanding of the 
phenomenon one needs to look for new CPV interactions;
especially those outside the $K^0$ system. It would be particularly
interesting if CP or T symmetry is also violated in flavor-conserving 
processes such as $\eta$ decays and the neutron electric dipole moment 
($d_n$) \cite{GN,Peter1}. The information here is very limited. Thus far 
the efforts have been
concentrated mainly in the search for the electric dipole moments (EDM) of the 
neutron and the electron
and a few nuclear reactions ( for a review see \cite{lk} ).
For EDM's, the experimental limit of $d_n<10^{-25}e\ cm$,
 which is far from the standard model prediction of $O(10^{-32})e\ cm$,
provides stringent constraints on various CP violating models beyond
the CKM paradigm. However, it is known that it will be very hard to 
improve the limit for another few orders due to limitations of  current 
technology.

 In this paper we wish to investigate the use of $\eta$ to decays as 
probes of CPV
in the flavor conserving sector. This is prompted  by the 
possibility of producing a large number of $\eta$ mesons
in  high statistics experiments. This new possibility can provide us with 
tools to gain more
knowledge on rare $\eta$ decay processes and offer unique  probes of
new physics. As we shall see the $\eta$ meson can yield complementary
information on strangeness conserving CPV which can only be indirectly 
gleamed from
neutron EDM studies.

To begin  we will study CPV effects in
\be
\eta(p)\to\pi^+(p_+)\pi^-(p_-)\gamma (k,\epsilon)\,,
\label{decay}
\ee
where $\epsilon$ is the photon polarization and obvious kinematics notation. 
This decay has a  
branching ratio of 4.75 percent and relatively simple final states.
Under the Lorentz and gauge invariances, the general amplitude
for the decay in Eq. (\ref{decay})
is given by
\be
{\cal M} &=& {i\over m_{\eta}^3}\{-M\varepsilon_{\mu\nu\rho\lambda}
p_+^{\mu}p_-^{\nu}k^{\rho}\epsilon^{\lambda}+
E[(\epsilon\cdot p_+)(k\cdot p_-)-(\epsilon\cdot p_-)(k\cdot p_+)]\}\,,
\label{Amp1}
\ee	
where the terms corresponding to $M$ and $E$ stand for magnetic and 
electric transitions, which are P-conserving and violating, respectively,
since $C(\eta)=+$ and $P(\eta)=-$. 
We note that since $C(\pi^+\pi^-\gamma)=(-1)^{l+1}$ with $l$ being
the angular momentum of the dipion, the 
$\pi^+\pi^-$ state has to be in an odd angular momentum state if
$C$ is conserved. 
The diagrams that contribute to the decay
in Eq. (\ref{decay}) are shown in Figure 1.
We note that the terms 
associated with $M$ and $E$ in 
Eq. (\ref{Amp1}) are given only up to third order in momenta and $M$ and 
$E$ are functions of Lorentz scalars.

In  the $\eta$ rest frame , without loss of generality we can choose 
the decay to be  
 in the $x-z$ plane . In this 
frame, the photon momentum can be directed  
along the $z$-axis and we denote its energy by $E_{\gamma}$.
The amplitude in Eq. (\ref{Amp1}) then becomes 
\be
{\cal M} &=& im_{\eta}^{-2}E_{\gamma}[M\hat{k}\cdot (\vec{\epsilon}\times 
\vec{p}_+)-E\vec{\epsilon}\cdot\vec{p}_+]\,.
\label{Amp2}
\ee	
If both $M$ and $E$ are $C$-even states, $i.e.$,
$M=M(s,(k\cdot q)^2)$ and $E=E(s,(k\cdot q)^2)$ with 
$s=(p_++p_-)^2=m_\eta^2-2m_\eta E_{\gamma}$ and $q=p_+-p_-$, 
the terms with $M$ and $E$ in Eq. 
(\ref{Amp1}) are (C, P, CP)=(+,-,-) and (+,+,+). 
In the absence of final state interactions,
$M$ and $E$ are purely real due to the $CPT$ theorem.
In this case, the existence of the $E$ term is clearly a indication of
direct CP violation since $CP(\eta)=-$.
We note that $M$ and $E$ are C-odd if
$M=m_\eta^{-2}\,k\cdot q\,M'(s,(k\cdot q)^2)$ and 
$E=m_\eta^{-2}\,k\cdot q\,E'(s,(k\cdot q)^2)$ \cite{TDLee,Peter2}.
The C violating interaction with $C(M)=-$ can be induced in models such as 
 noncommutative QED associated with studies in string theory  \cite{NCFT} 
through $\eta\to 3\gamma$ \cite{3gamma}. However, the effect is 
negligibly small and experimentally it is consistent with zero
\cite{PDG01}.

In general, to observe a CP violating effect, one needs to study the 
interference between the $M$ and $E$ terms with explicit photon polarization.
To show this, we write the squared amplitude from Eq.(\ref{Amp2}) as
\be
|{\cal M}|^2 &=&
m_{\eta}^{-4}E_{\gamma}^2\left\{ 
|M|^2|\hat{k}\cdot (\vec{\epsilon}\times 
\vec{p}_+)|^2+|E|^2|\vec{\epsilon}\cdot\vec{p}_+|^2+
\right.
\nn\\
&&
\left.
 E^*M[\hat{k}\cdot (\vec{p}_+\times\vec{\epsilon})]
\,(\vec{\epsilon}\cdot\vec{p}_+)^*+
M^*E[\hat{k}\cdot (\vec{p}_+\times\vec{\epsilon})]^*
\,(\vec{\epsilon}\cdot\vec{p}_+)\right\}\,.
\label{Amp3}
\ee
It is easily seen that the 
interference terms 
in Eq. (\ref{Amp3}) are related to the triple momentum correlation of
$\vec{k}\cdot (\vec{p}_+\times\vec{\epsilon})$, 
which is odd under the time-reversal transformation.
Observing this correlation would  be a 
 sign of direct T violation.
By summing over photon polarizations, from 
Eq. (\ref{Amp3}) the partial decay rate
of the Dalitz plot density is
\be
\frac{d \Gamma}{dE_+ \, dE_-} & = & \frac{1}{64 \pi^3m_{\eta}}
\sum_{spin}|{\cal M}|^2\;\propto\;
 |E|^2 + |M|^2 \,,
\label{dGamma1}
\ee
where $E_{\pm}$ are the $\pi^{\pm}$ energies. 
In terms of $E_{\gamma}$, it can be also written as
\be
\frac{d \Gamma}{dE_{\gamma}\, d\cos\theta} & = & \frac{1}{512 \pi^3}
\left( \frac{E_{\gamma}}{m_{\eta}} \right)^3 \beta^3 \left( 1- \frac{2
E_{\gamma}}{m_{\eta}}\right) \sin^2 \theta
\left[ |E|^2 + |M|^2 \right]\,,
\label{dGamma2}
\ee
where $\beta=(1-4m_{\pi}^2/s)^{1/2}$ and $\theta$
is the angle between $\pi^+$ and $\gamma$ in the $\pi^+ \pi^-$ rest frame.
As seen from Eqs. ({\ref{dGamma1}) and ({\ref{dGamma2}),
there is no interference term between $M$ and $E$, and 
therefore, no CP or T violation can be detected  without the explicit measurement of the photon 
polarization.

 From Eq. (\ref{Amp3}), 
we can  define the photon polarization in terms of the well 
known density matrix\footnote{The analysis for the photon polarization in 
the case of $K_L\to\pi^+\pi^-\gamma$ can be found in Ref. 
\cite{Sehgal}.}
 \begin{equation}   
\rho = \left( \begin{array}{cc} |E|^2 & E^*M \\ EM^* & |M|^2 \end{array}
\right) = \frac{1}{2} \left( |E|^2+|M|^2 \right) \left[ 1 \!\!\!\!\:\: 
\mbox{l}
 + \vec{S}(E_\gamma,\theta) \cdot \vec{\tau}  \right]
\end{equation}
where $\vec{\tau} = (\tau_1, \, \tau_2, \, \tau_3)$ denotes the Pauli 
matrices,
and $\vec{S}$ is the Stokes vector of the photon with components
\begin{eqnarray}
S_1(E_\gamma,\theta) & = & 2 Re \left( E^*M \right) / \left( |E|^2 + |M|^2 
\right)\,,
\nn\\
S_2(E_\gamma,\theta) & = & 2 Im \left( E^*M \right) / \left( |E|^2 + |M|^2 
\right)\,,
\nn\\ 
S_3 (E_\gamma,\theta)& = & \left(|E|^2 - |M|^2 \right) / \left( |E|^2 + 
|M|^2 \right)\,,
\label{Stokes}
\end{eqnarray}
respectively.
Here, we have used 
\be
\vec{\epsilon}_1&=& {(\vec{p}_+\times \vec{k})\times \vec{k}\over
|(\vec{p}_+\times \vec{k})\times \vec{k}|}\,,
\nn\\
\vec{\epsilon}_2&=& {\vec{p}_+\times \vec{k}\over |\vec{p}_+\times 
\vec{k}|}\,,
\ee
as the two independent polarization vectors.
The Stokes parameters $S_{1,2}$ in Eq. (\ref{Stokes})
can be expressed as
\be
S_{1(2)} (E_\gamma,\theta) &=& 
{d\Gamma_{+(L)}-d\Gamma_{-(R)}
\over 
d\Gamma_{+(L)}+d\Gamma_{-(R)}}\,,
\label{PS12}
\ee
where $d\Gamma_{\pm}$ and
$d\Gamma_{L,R}$ refer to photons
polarized in the directions
\be
\vec{\epsilon}_{\pm} &=& {1\over 
\sqrt{2}}\left(\vec{\epsilon}_2\pm\vec{\epsilon}_1\right)\,,
\ee
which is $45^0$ with respect to the decay plane,
and
\be
\vec{\epsilon}_{L,R} &=& {1\over
\sqrt{2}}\left(1,\mp i,0\right)\,,
\ee
respectively. 
We can also define the integrated Stokes parameters $S_i(E_\gamma)$, for 
example, by integrating $\theta$ in Eq. (\ref{PS12}) for each 
$d\Gamma_\alpha\ (\alpha=\pm,L,R)$.
The two parameters of $S_{1,2}$  are sometimes called 
as
the {\em linear} and {\em circular} polarizations.
It is clear that
$S_1$ and $S_2$ are related to CPV.
 If $E=0$, one has that $S_1=S_2=0$ and $S_3=-1$ and there is no CP 
violation.

Phenomenologically, the decay rate of $\eta\to\pi^+\pi^-\gamma$ is 
described by the magnetic term from the box-anomaly and resonance 
contributions \cite{Eta}. Explicitly, for example,
in a chiral model it was found that \cite{Etaexpt,Picciotto,Rev}
\be
M &=& M^+(s)\;\simeq\; -{em_{\eta}^3\over 4\pi^2f_{\pi}^3}\times 
{\sqrt{3}\over 6}\left(
1-{3m_{\rho}^2\over m_{\rho}^2-s}\right)\,,
\label{Mchiral}
\ee
where $C(M^+)=+$, $f_{\pi}=93\ MeV$ and  $m_{\rho}=770\ MeV$.
 From Eq. (\ref{Mchiral}), we see that the dominant contribution to 
the decay is from a C-even state.
If we include terms that are higher order in momentum and thus nonleading
we have
\be
M &=& M^+ +M^-\,,
\nn\\
E &=& E^+ +E^-\,,
\label{EM}
\ee
where $C(M^{\pm},E^{\pm})=\pm$.
With final state interactions, in Table 1, we show the possible violations 
of symmetries for the photon polarizations $S_i$ with $M$ and $E$ given in 
Eq. (\ref{EM}).

\begin{table}[h]
   \caption{ Possible violations of symmetries for
the photon polarizations $S_i$ due to the interferences between $M^+$
and $M^-,E^\pm$ with final state interactions.}  
   \begin{center}
   \begin{tabular}{|c|c|c|c|}
   \hline
 Interference &  $S_1$ & $S_2$ & $S_3$  \\ \hline
$M^+M^-$ & $-$ & $-$ & $C$, $CP$   \\ \hline
$M^+E^+$ & $P$, $CP$  & $P$, $CP$ & $-$ \\ \hline
$M^+E^-$ & $C$, $P$  & $C$, $P$ & $-$\\ \hline
   \end{tabular}
   \end{center}
\end{table}

\begin{table}[h]
   \caption{ 
Same as Table 1 but without final state phases.}
   \begin{center}
   \begin{tabular}{|c|c|c|c|}
   \hline
 Interference &  $S_1$ & $S_2$ & $S_3$  \\ \hline
$M^+M^-$ & $-$ & $-$ &  $-$   \\ \hline
$M^+E^+$ & $P$, $CP$  &  $-$ & $-$ \\ \hline
$M^+E^-$ & $-$  & $C$, $P$ & $-$\\ \hline
   \end{tabular}
   \end{center}
\end{table}

We now study the possible interactions which would yield the electric 
transitions in Eq. (\ref{Amp1}).
First of all, it is easy to see that the $E$ term can be induced first going
through the  $ \pi^+\pi^-$ intermediate state which
violates CP symmetry because $CP(\pi^+\pi^-)=+$, and then radiating the photon
from the pions. These bremsstrahlung terms are shown in Figures 1b and 1c. 
Explicitly, we have that
\be
E^+(\eta\to(\pi^+\pi^-)^*\to \pi^+\pi^-\gamma)
&=& {em_{\eta}^3g_{\eta\pi\pi}\over (p_+\cdot k)(p_-\cdot k)}\,,
\label{Eppg}
\ee
where $g_{\eta\pi\pi}$ is the effective coupling for $\eta\to \pi^+\pi^-$.
 From the experimental limit of $Br(\eta\to\pi^+\pi^-)<3.3\times 10^{-4}$, 
we find that
\be
|g_{\eta\pi\pi}|^{exp}&<& 1.2\times 10^{-4}\ GeV\,.
\label{gepp}
\ee
To illustrate the order of limits on the CP violating effects,
in Figure 2,  by using
Eqs. (\ref{PS12}), (\ref{Mchiral}) and (\ref{Eppg}), 
we show $S_1(E_\gamma)$  with
a real upper value of $g_{\eta\pi\pi}$ in Eq. (\ref{gepp}).
However, if we assume that the relative strong phase\footnote{
The dipion state for the $M^+$ transition in Eq. (\ref{Mchiral})
is in a state of angular momentum $l_{\pi\pi}=1$ and isospin $I=1$,
while that in Eq. (\ref{Eppg}) is $l_{\pi\pi}=0$ and isospin $I=0$,
with the final state phase shifts being $\delta^1_1$ and $\delta^0_0$,
respectively, and $\delta=\delta^1_1-\delta^0_0$.}  
between the terms of $M^+$ and 
$E^+$
is $\delta$, $i.e.$, $g_{\eta}=|g_{\eta}|e^{i\delta}$, 
 from Figure 2,
we get that
\be
|S_{1,2}(E_\gamma)|&<& 0.2\cos\delta\,,\  0.2\sin\delta \,,
\ \ \ {\rm and}\ S_3\simeq -1\,,
\label{Spp}
\ee
for $E_\gamma>20\ MeV$. It is interesting to see from
Eq. (\ref{Spp})
 that one can get rid of the strong phase by measuring
$S_1^2+S_2^2$ to give the pure CP violating effect.\footnote{We note that
$S_1^2+S_2^2=1-S_3^2$ which is zero if there is no CP violation.}

Within the Standard model the sources for the decay of
$\eta\to\pi^+\pi^-$ can arise 
from the CKM phase, 
and/or the strong $\theta$ term in QCD \cite{theta,Jarlskog1}. 
New physics such as  spontaneous CP violating models can 
also lead to this decay \cite{Jarlskog2}.
 From the experimental data on CP violation, 
it is found that \cite{Jarlskog1,Jarlskog2}
 $|g_{\eta\pi\pi}|$ are less than
$2.6\times 10^{-16}$, $2\times 10^{-10}$, and $5\times 10^{-11}\ GeV$, 
for the above three sources, respectively.
In these cases, the CP violating effect such as the parameters of 
$|S_{1,2}|$ are expected to be 
less than $2.2\times 10^{-12}$, $1.7\times 10^{-6}$ and 
$4.2\times 10^{-7}$, respectively,
which are vanishingly small and thus undetectable.
This is not surprising at all since the constraints from 
the CP violating parameters from  $\epsilon$ and $\epsilon'$ 
in the $K^0$-system as well
as $d_n$ are very strong in these  models.
To evade these constraints, one has to search for some 
unconventional sources of CPV which do not contribute directly
to $\epsilon$, $\epsilon'$ and $d_n$ and yet has a contribution in
$\eta\to\pi^+\pi^-\gamma$. This leads us to search for operators that do 
not
contribute directly to
$\eta\to\pi^+\pi^-$ and the well studied $K^0$ decays.
Hence, we  construct
flavor-conserving CP violating four-fermion operators involving two strange
quarks together with combinations of other light quarks.

 Explicitly, we study the four-fermion operator, 
given by
\be
{\cal O} &=& {1\over m_{\eta}^3}G\,\bar{s}
i\sigma_{\mu\nu}\gamma_5(p-k)^{\nu}\,s\,
\bar{u}\gamma^{\mu}u\,,
\label{O}
\ee
where $u (s)$ stands for the up (strange) quark and
$G$ is a dimensionless parameter originating from 
short distance 
physics and it can
be taken real due to the CPT theorem and taking $C(G)=+$.
 We note that there is no charge asymmetry in the decay due to the $C$ 
invariance and we also note that
one may discuss similar operators
with appropriate color generators and indices included
in Eq. (\ref{O}).

To calculate its contribution to $\eta\to\pi^+\pi^-\gamma$,
we use a factorization approximation that the $\pi^+\pi^-$ part is from 
$\bar{q}\gamma^{\mu}q$ and
the $\eta \gamma$ transition involves only part containing strangeness,
$i.e.$,
\be
<\eta|{\cal O}|\pi^+\pi^-\gamma>
&\sim & {1\over m_\eta^3}G
<\eta|\bar{s}i\sigma_{\mu\nu}\gamma_5(p-k)^{\nu}\,s|\gamma>
<0|\bar{u}\gamma^\mu\,u|\pi^+\pi^->\,.
\ee
This can be viewed as the photon being emitted directly from
the structure part shown in Figure 3. 
We define the form factors for the $\eta\to \gamma$ transition by
\be
<\eta|\bar{s}i\sigma_{\mu\nu}\gamma_5(p-k)^{\nu}\,s|\gamma>
&=&ie\left[\epsilon_{\mu}(k\cdot p)-(\epsilon\cdot p)k_{\mu}\right]
{F(s)\over m_{\eta}}\,.
\label{FF}
\ee
The form factor of $F(s)$ can be estimated in various
QCD models such as the light front quark model (LFQM) \cite{LFQM}
and the relativistic  quark model (QM) \cite{QM}.
In the LFQM \cite{LFQM}, 
we find that, for example,  $F(0)\sim 0.19$ which is consistent with
the result from the QM \cite{QM}. We shall assume that $F(s)\sim F(0)$
to illustrate our numerical values for CP violation.

 From Eqs. (\ref{O}) and (\ref{FF}), we obtain
that
\be
E&\sim & 2eF(s)G\,.
\label{Ei}
\ee
For $G<O(1)$, we have that $|M|>>|E|$.
 From Eqs. (\ref{PS12}), (\ref{Mchiral}) and (\ref{Ei}),
we find that
\be
S_1(E_\gamma,\theta)&=&S_1(E_\gamma)\sim \frac{96\pi^2F(s)G 
}{\sqrt{3}}\left({f_\pi\over m_\eta}\right)^3
\left(\frac{ m_\rho^2-s }{2m_\rho^2+s} \right )\,,
\nn\\
S_2 &\sim & 0\,,
\ \  {\rm and} \ \ S_3\;\sim \; -1\,,
\label{NS123}
\ee
where we have assumed that $C(G)=+$, $i.e.$, $E\sim E^+$.
The results in Eq. (\ref{NS123}) indicate that the only interesting
CPV observable is the linear polarization $S_1$
unless $G$ are C-odd, such as when they contain higher order 
momentum terms, i.e.
 $G\propto k\cdot q$.
But in this case the circular
polarization of $S_2$ which can be non-zero are both CP and T 
conserving as shown in Table 2 and 
therefore we shall not discuss it in the remaining of
the paper.
 In Figure 4, we show the linear polarization of the Stokes parameter 
$S_1$ as a function of $E_\gamma=(m_\eta^2-s)/2m_\eta$ with $G\sim 1$. 
For the maximal energy of the photon, $i.e.$, $E_\gamma\simeq 0.2\ GeV$, 
$S_1(E_\gamma)$ is about $21\%$. 

We now discuss 
possible constraints on $G$ and
 models that may lead to the operator in 
Eq. (\ref{O}).
As we mentioned previously, the operator in Eq. (\ref{O})
 cannot directly generate the decay of $\eta\to\pi^+\pi^-$.
It is also important to note that it cannot dircetly induce $d_n$ either;
unlike the scalar-type operators studied in Ref. \cite{Pospelov}.
There are basically no direct  constraints 
for $G$ from both low and high energy experiments.
In principle, $G\sim O(1)$ is allowed. A search for the photon 
polarization in $\eta\to\pi^+\pi^-\gamma$ will therefore provide 
interesting limits on $G$
in the event of a negative search.

Our study have focused on the general aspects of $CPV$ in Eq.(\ref{decay})
and appeal to experimantal bounds as much as possible without committing
to models explicit of $CP$ violation. 
However, one would be wondering how the operator in Eq. (\ref{O})
 can be realized from some theoretical models.

In the standard model, there is no contribution to 
the operator in Eq. (\ref{O}) at the lowest order and also to 1-loop.
However, the operator can be 
induced in models in which the strange quark has an EDM. But, it has 
to be generated from a weak loop in a CP violating theory as well. Details are
highly model dependent; however, we can expect that $G$ will not be larger than
$G_Fm_\eta^2\simeq 0.35\times 10^{-5}$. This leads to the expectation
that $S_1$ is less than $10^{-6}$.

Finally, we remark that the operator given by the replacement of 
$s$ by $u$ or $d$ in Eq. (\ref{O}) can also induce $S_1$ but
it is very small due to the constraint from the neutron EDM.\\

\noindent {\bf Acknowledgments}

This work was supported in part by the National Science Council of the
Republic of China under Contract No. NSC-90-2112-M-007-040 and National 
Science and Engineering Research Council of Canada.

%
%
%



\newpage
\begin{figure}[h]
\includegraphics{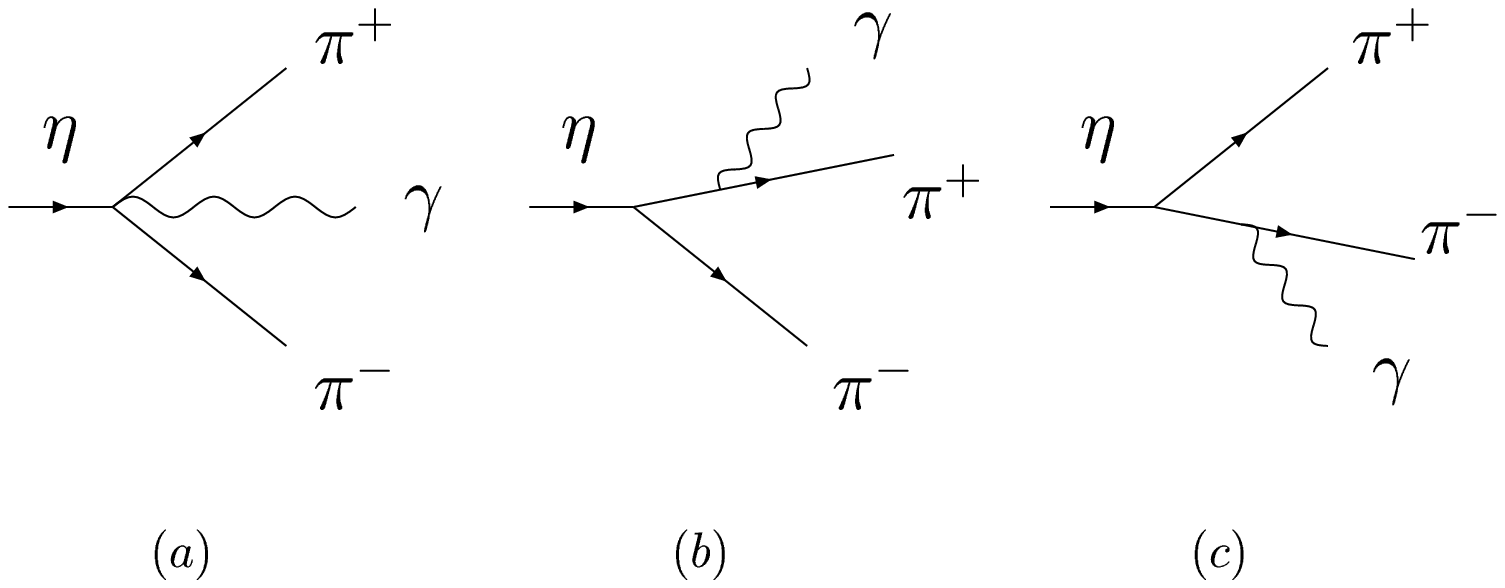} 
\vskip 5.cm \caption{
Diagrams which contribute to $\eta\to\pi^+\pi^-\gamma$.}
 \label{Feynman}
\end{figure}

\vskip 1.cm
\begin{figure}[h]
\includegraphics{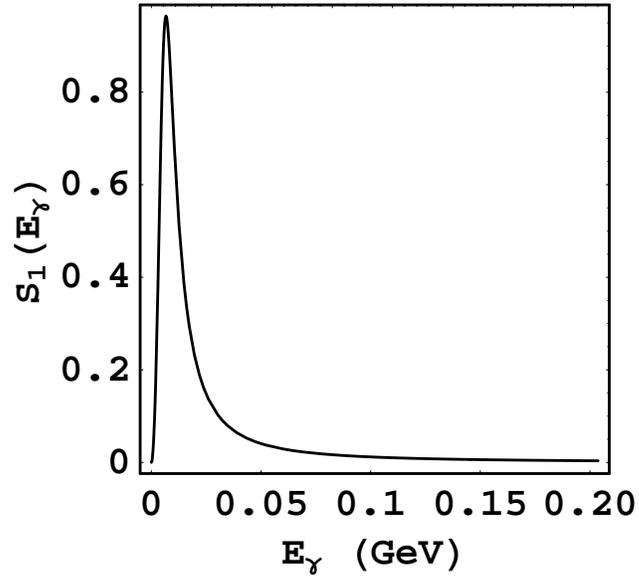} 
\vskip 9.cm 
\caption{ 
The Stokes parameter $S_1(E_\gamma)$
as a function of
 $E_\gamma$ due to $\eta\to\pi^+\pi^-$.}
\end{figure}

\begin{figure}[h]
\includegraphics{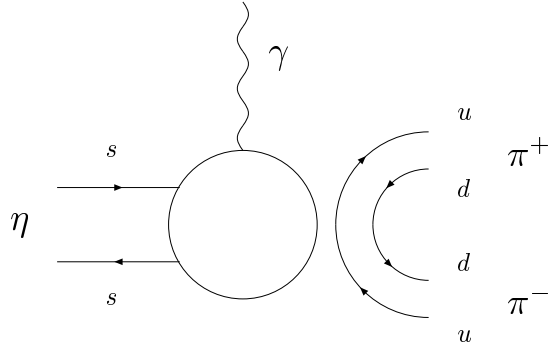} 
\vskip 7.cm 
\caption{ 
Structure dependent contribution to $\eta\to\pi^+\pi^-\gamma$.}
\end{figure}

\begin{figure}[h]
\includegraphics{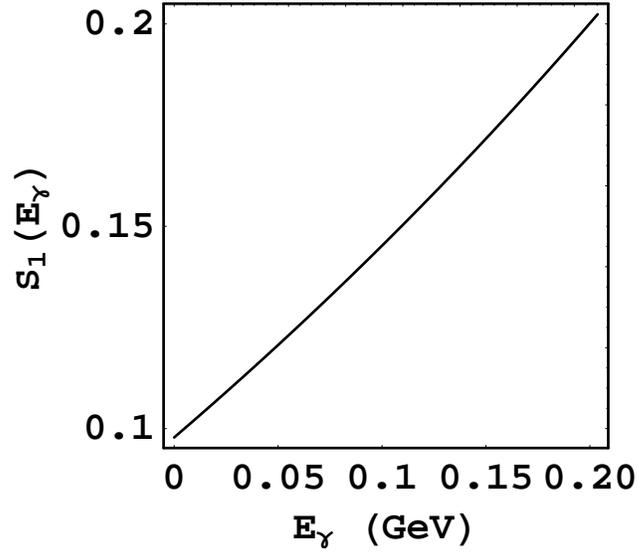} 
\vskip 9cm 
\caption{ 
The Stokes parameter $S_1(E_\gamma)$
as a function of
 $E_\gamma$ due to the operator in Eq. (\ref{O}).}
\end{figure}

\ed